\documentclass[prd,12pt]{revtex4}
\pdfoutput=1 
\usepackage{amsmath}
\usepackage{graphicx}
\usepackage{subfigure}
\usepackage{bm}
\usepackage{slashed}
\usepackage{lineno}

\def\d{\mathrm{d}}
\def\beq{\begin{equation}}
\def\eeq{\end{equation}}
\def\bea{\begin{eqnarray}}
\def\eea{\end{eqnarray}}

\def\ba{\begin{array}}
\def\ea{\end{array}}

\DeclareMathOperator{\Li}{Li}

\begin{document}
\title{The isospin asymmetry in $B \to K^* \mu^+ \mu^-$ using AdS/QCD}

\author{M. Ahmady}
\affiliation{Department of Physics, Mount Allison University, Sackville, N-B. E46 1E6, Canada}
\email{mahmady@mta.ca} 


\author{S. Lord}
\affiliation{D\'epartement de Math\'ematiques et Statistique, Universit\'e de Moncton,
Moncton, N-B. E1A 3E9, Canada}
\email{esl8420@umoncton.ca}

\author{R. Sandapen}
\affiliation{D\'epartement de Physique et d'Astronomie, Universit\'e de Moncton,
Moncton, N-B. E1A 3E9, Canada
\& \\
Department of Physics, Mount Allison University, Sackville, N-B. E46 1E6, Canada }
\email{ruben.sandapen@umoncton.ca} 

\begin{abstract}
We compute the isospin asymmetry distribution in the rare dileptonic decay $B \to K^* \mu^+ \mu^-$, in the dimuon mass squared ($q^2$) region below the $J/\Psi$ resonance, using non-perturbative inputs as predicted by the anti-de Sitter/Quantum Chromodynamics (AdS/QCD) correspondence and by  Sum Rules. We predict a positive asymmetry at $q^2=0$ which flips sign in the region  $q^2 \in [1,2]~\mbox{GeV}^2$  to remain small ($\le 2\%$) and negative for larger $q^2$. While our predictions are distinct as $q^2 \to 0$, they become hardly model-dependent $q^2 \ge 4~\mbox{GeV}^2$. We compare our predictions to the most recent LHCb data. 
\end{abstract}


\maketitle

\section{Introduction}
The rare decay $B \to K^* \mu^+ \mu^-$ has recently been attracting much attention  from both the experimental \cite{Aaij:2014pli,Aaltonen:2011ja,Lees:2012tva,Wei:2009zv,Aaij:2013qta,Aaij:2012cq,Aaij:2013iag,LHCb-CONF-2012-008} and theoretical \cite{Altmannshofer:2013foa,Descotes-Genon:2013wba,Hurth:2013ssa,Descotes-Genon:2013vna,Gauld:2013qba,Buras:2013qja,Gauld:2013qja,Hambrock:2013zya,Khodjamirian:2010vf,Bharucha:2010im,Datta:2013kja,Alok:2009tz} sides because various observables associated with this decay are susceptible to reveal New Physics effects. An interesting observable to look at is the isospin asymmetry distribution defined as 
\begin{equation}
A_I (q^2)=\frac{\d \Gamma(B^0 \to K^{*0} \mu^+ \mu^-)/\d q^2-\d \Gamma(B^+ \to K^{*+} \mu^+ \mu^-)/\d q^2}{\d \Gamma(B^0 \to K^{*0} \mu^+ \mu^-)/\d q^2+ \d \Gamma(B^+ \to K^{*+} \mu^+ \mu^-)/\d q^2}
\end{equation}
since being a ratio of differential decay widths, the leading uncertainties in the $B \to K^*$ form factors cancel in the theoretical computation of this asymmetry. Nevertheless, there remains some model-dependence in theory predictions which we address in this paper.

In a previous paper \cite{Ahmady:2013cva}, two of us have computed the isospin asymmetry in $B \to K^* \gamma$ where we highlighted that an advantage of using an AdS/QCD twist-$2$ DA is that it avoids the end-point divergence encountered with the corresponding Sum Rules DA. We now extend our calculation for  $B \to K^* l^+ l^-$, i.e. for the case $q^2 \ne 0$, where $q^2$ is the dimuon mass squared. This isospin aymmetry distribution has recently been measured by the LHCb Collaboration at $3~\mbox{fb}^{-1}$ \cite{Aaij:2014pli} superseding  the previous LHCb  measurements at $1~\mbox{fb}^{-1}$ given in Ref.\cite{Aaij:2012cq}. The original SM computation of the isospin asymmetry in $B \to K^* \mu^+ \mu^-$ was performed by Feldmann and Matias in Ref. \cite{Feldmann:2002iw} which we follow here. A more sophisticated calculation of this isospin asymmetry has recently been performed by Lyon and Zwicky in Ref. \cite{Lyon:2013gba}.

A potential source of theoretical uncertainty in the SM prediction arises from the model-dependent non perturbative quantities, namely the Distribution Amplitudes and decay constants of the $K^*$ as well as the two universal soft $B \to K^*$ transition form factors. The latter form factors are deduced from the seven $B \to K^*$ form factors which themselves can be obtained from lattice QCD at high $q^2$ \cite{Horgan:2013hoa} and from  light-cone sum rules (LCSR) at low to moderate $q^2$ \cite{Aliev:1996hb} .  LCSR require as non-perturbative inputs the model-dependent DAs of the $K^*$ meson as well as its decay constants.

Our goal in this paper is to repeat the computation of Ref. \cite{Feldmann:2002iw} for the isospin asymmetry distribution in $B \to K^* l^+ l^-$ but with different inputs for the non-perturbative quantities mentioned above. We compute the decay constants and the DAs of the $K^*$ using AdS/QCD \cite{Ahmady:2013cva} while we build upon our previous work \cite{Ahmady:2014sva}  
to obtain the two universal $B \to K^*$ soft form factors. To investigate the degree of model-dependence, we shall compare our AdS/QCD prediction to that obtained using Sum Rules DAs and decay constants.

\section{The isospin asymmetry}
In the QCD factorization approach, the isospin asymmetry distribution is given by \cite{Feldmann:2002iw}
\begin{equation}
A_I(q^2)=\Re e (b_d^{\perp}(q^2)-b_u^{\perp}(q^2)) \frac{|\mathcal{C}_9^{(0)\perp}(q^2)|^2}{|C_{10}(\mu_b)|^2 +|\mathcal{C}_9^{(0)\perp}(q^2)|^2 } \times \frac{F(q^2)}{G(q^2)}
\label{isoasy-th}
\end{equation}
with
\begin{equation}
F(q^2)=1 + \frac{1}{4} \frac{E_{K^*}^2m_B^2}{q^2m_{K^*}^2} \frac{\xi_{\parallel}^2(q^2)}{\xi_{\perp}^2(q^2)} \frac{\Re e (b_d^{\parallel}(q^2)-b_u^{\parallel}(q^2)) }{\Re e (b_d^{\perp}(q^2)-b_u^{\perp}(q^2)) } \frac{|\mathcal{C}_9^{(0)\parallel}(q^2)|^2}{|\mathcal{C}_9^{(0)\perp}(q^2)|^2 }
\label{Fq2}
\end{equation}
and
\begin{equation}
G(q^2)=1 + \frac{1}{4} \frac{E_{K^*}^2m_B^2}{q^2m_{K^*}^2} \frac{\xi_{\parallel}^2(q^2)}{\xi_{\perp}^2(q^2)} \frac{|\mathcal{C}_9^{(0)\parallel}(q^2)|^2 +|C_{10}(\mu_b)|^2 }{|\mathcal{C}_9^{(0)\perp}(q^2)|^2 + |C_{10}(\mu_b)|^2} 
\label{Gq2}
\end{equation}
where $E_{K^*}=(m_B^2-q^2)/(2m_B)$ is the energy of the $K^*$ meson. The generalized SM Wilson coefficients are given by 
\begin{equation}
C_9^{(0)\perp}(q^2)=C_9(\mu_b) + Y(q^2) + \frac{2m_b m_B}{q^2} C_7^{\rm eff} (\mu_b)
\label{C90perp}
\end{equation}
and
\begin{equation}
C_9^{(0)\parallel}(q^2)=C_9(\mu_b) + Y(q^2) + \frac{2m_b}{m_B} C_7^{\rm eff}(\mu_b) \;.
\label{C90parallel}
\end{equation}
As noted in Ref. \cite{Feldmann:2002iw}, in the limit $q^2 \to 0$, the photon pole in $C_9^{(0) \perp}$ dominates and  Eqn. \eqref{isoasy-th} becomes
\begin{equation}
A_I(0)=\Re e (b_d^{\perp}(0)-b_u^{\perp}(0)) 
\end{equation}
which is  the isospin asymmetry in $B \to K^* \gamma$ computed originally in Ref. \cite{Kagan:2001zk}.   In the definitions \eqref{C90perp} and \eqref{C90parallel}, the function $Y(q^2)$ is given by \cite{Beneke:2001at} 
\begin{eqnarray}
\label{yy}
Y(q^2) &=& h(q^2,m_c,\mu_b) \left(3 \overline{C}_1(\mu_b) +\overline{C}_2(\mu_b)+3 
\overline{C}_3(\mu_b)+\overline{C}_4(\mu_b) +3 \overline{C}_5(\mu_b)+\overline{C}_6(\mu_b)\right) 
\nonumber\\
&-&\,\frac{1}{2}\,h(q^2,m_b,\mu_b) \left(4 \,(\overline{C}_3(\mu_b)+\overline{C}_4(\mu_b))+3 
\overline{C}_5(\mu_b)+\overline{C}_6(\mu_b)\right) 
\nonumber\\
&-&\frac{1}{2}\,h(q^2,0,\mu_b) \left(\overline{C}_3(\mu_b)+3 \overline{C}_4(\mu_b)\right)+\,\frac{2}{9}\,\left(\frac{2}{3}\overline{C}_3(\mu_b)+2 \overline{C}_4(\mu_b)+\frac{16}{3}  
\overline{C}_5(\mu_b)\right)
\end{eqnarray}
where the vacuum polarisation function
\begin{equation}
h(q^2,m_q,\mu_b)=-\frac{4}{9} \left(6 \int_0^1 x(1-x) \ln (m_q^2-x(1-x)q^2-i \epsilon) \d x -\ln(\mu_b^2) + 1 \right) \;.
\end{equation}
In Eqns. \eqref{isoasy-th} and \eqref{Fq2}, 
\begin{equation}
b_q^{\perp}(q^2)=\frac{24 \pi^2 m_B f_B e_q}{q^2 \xi_{\perp}(q^2) \mathcal{C}_9^{(0)\perp}(q^2)} \left(  \frac{f_{K^*}^\perp}{m_b} K_1^{\perp} (q^2) + \frac{f_K^* m_K^*}{6 \lambda_{B,+}(q^2) m_B} \frac{K_2^{\perp} (q^2)}{1-q^2/m_B^2}\right)
\label{bperpq2}
\end{equation}
and
\begin{equation}
b_q^{\parallel}(q^2)=\frac{24 \pi^2 f_B e_q m_{K^*}}{m_B E_{K^*} \xi_{\parallel}(q^2) \mathcal{C}_9^{(0)\parallel}(q^2)} \left( \frac{f_{K^*}}{3\lambda_{B,-} (q^2)} K_1^{\parallel}(q^2) \right)  
\label{bparallelq2}
\end{equation}
where we take $f_B=0.190$ GeV. \footnote{We infer this value from the branching ratio datum for $B^+ \to \tau^+ \nu_{\tau}$ \cite{Beringer:1900zz}.} In Eqn. \eqref{bperpq2},
\begin{equation}
K_1^{\perp}(q^2)=K_1^{\perp(a)}(q^2)+K_1^{\perp(b)}(q^2)+K_1^{\perp(c)}(q^2)
\end{equation}
with
\begin{equation}
K_1^{\perp(a)}(q^2)=-\left( \overline{C}_6(\mu_h) + \frac{\overline{C}_5(\mu_h)}{N_c} \right) F_{\perp}(\hat{s}),  \hspace{0.5cm} F_{\perp}(\hat{s}) =\frac{1}{3} \int_0^1 \d u \frac{\phi_{\perp} (u)}{\bar{u} + u \hat{s}} \;,
\label{K1perpa}
\end{equation}
\begin{equation}
 K_1^{\perp(b)}(q^2)=C_8^{\rm eff} (\mu_h) \frac{m_b^2}{m_B^2} \frac{C_F}{N_c} \frac{\alpha_s(\mu_h)}{4 \pi} X_{\perp}(\hat{s}), \hspace{0.5cm} X_{\perp}(\hat{s})=F_{\perp}(\hat{s}) + \frac{1}{3} \int_0^1 \d u \frac{\phi_{K^*}^{\perp}(u)}{(\bar{u}+ u \hat{s})^2}  \;,
\label{K1perpb}
\end{equation}
\begin{equation}
K_1^{\perp(c)}(q^2)= \frac{C_F}{N_c}\frac{\alpha_s(\mu_h)}{4\pi} \frac{2}{3} \int_0^1 \d u \frac{\phi_{K^*}^{\perp} (u)}{\bar{u} + u \hat{s}} F_V(\bar{u} m_b^2 + u q^2) \;
\label{K1perpc}
\end{equation}
and
\begin{equation}
K_2^{\perp}(q^2)=K_2^{\perp(a)}(q^2)+K_2^{\perp(b)}(q^2)+K_2^{\perp(c)}(q^2)
\end{equation}
with
\begin{equation}
K_2^{\perp(a)}(q^2) = -\frac{\lambda_u}{\lambda_t}\left( \frac{\overline{C}_1}{3} (\mu_h)+ \overline{C}_2(\mu_h) \right)\delta_{qu} + \left( \overline{C}_4 (\mu_h)+ \frac{\overline{C}_3(\mu_h)}{3} \right) \;,
\label{K2perpa}
\end{equation}
\begin{equation}
K_2^{\perp(b)}(q^2) = \mathcal{O}\left( \frac{\Lambda_h}{m_B} \right) \;,
\label{K2perpb}
\end{equation}
\begin{equation}
K_2^{\perp(c)}(q^2)=-\frac{C_F}{N_c}\frac{\alpha_s(\mu_h)}{4\pi} 2 \int_0^1 \d u \left( g_\perp^{(v)}(u) - \frac{ {g'}_\perp^{(a)}(u)}{4} \right) F_V(\bar{u} m_b^2 + u q^2) \;.
\label{K2perpc}
\end{equation}
In Eqn. \eqref{bparallelq2},
\begin{equation}
K_1^{\parallel}(q^2)=K_1^{\parallel(a)}(q^2)+K_1^{\parallel(b)}(q^2)+K_1^{\parallel(c)}(q^2)
\end{equation}
with
\begin{equation}
K_1^{\parallel(a)}(q^2) = K_2^{\perp(a)}(q^2)
\label{K1parallela}
\end{equation}
\begin{equation}
K_1^{\parallel(b)}(q^2) = -C_8^{\rm eff}(\mu_h) \frac{m_b}{m_B} \frac{C_F}{N_c} \frac{\alpha_s(\mu_h)}{4\pi} F_\parallel(\hat{s}), \hspace{0.5cm} F_{\parallel}(\hat{s}) =2 \int_0^1 \d u \frac{\phi_{\parallel} (u)}{\bar{u} + u \hat{s}} \;,
\label{K1parallelb}
\end{equation}
\begin{equation}
K_1^{\parallel(c)}(q^2) = - \frac{C_F}{N_c}\frac{\alpha_s(\mu_h)}{4\pi}2 \int_0^1 \d u \phi_\parallel(u)F_V(\bar{u} m_b^2 + u q^2) \;. 
\label{K1parallelc}
\end{equation}
The vector form factor $F_V(s)$ appearing in Eqns \eqref{K1perpc}, \eqref{K2perpc},  and \eqref{K1parallelc}   is given as \cite{Feldmann:2002iw} 
\begin{eqnarray}
F_V(s) &=&\frac{3}{4} \{h(s,m_c,\mu_h) \,(\overline{C}_2(\mu_h) +\overline{C}_4(\mu_h)+\overline{C}_6(\mu_h))  \\ \nonumber
&+&h(s,m_b, \mu_h)\,(\overline{C}_3(\mu_h)+\overline{C}_4(\mu_h)+\overline{C}_6(\mu_h))\\ \nonumber
&+&h(s,0, \mu_h)\, (\overline{C}_3(\mu_h)+3 \overline{C}_4(\mu_h)+3 \overline{C}_6(\mu_h)) 
-\frac{8}{27}\,(\overline{C}_3(\mu_h)-\overline{C}_5(\mu_h)-15\overline{C}_6(\mu_h)) \}
\end{eqnarray}
Finally, in Eqns \eqref{bperpq2} and \eqref{bparallelq2},
\begin{equation}
\lambda_{B,\pm}^{-1}(q^2) =
  \int_0^\infty d\omega \, \frac{\phi^B_\pm(\omega)}{\omega-q^2/m_B - i
    \epsilon} \ .
\end{equation}
where $\phi^B_\pm(\omega)$ are given in Ref. \cite{Beneke:2000wa}.

Note that some of the above equations differ  from those given in Ref. \cite{Feldmann:2002iw}.   First, we take the argument of $F_V$ to be $\bar{u} m_b^2 + u q^2$ instead of $\bar{u} m_B^2 + u q^2$. Secondly, in Eq. \eqref{bperpq2}, we write 
$(f_{K^{*}}^{\perp}/m_b)$ instead of $(f_{K^{*}}^{\perp}/m_B)$. Thirdly, in Eqn. \eqref{K1perpb}, we have $(m_b^2/m_B^2)$ instead of $(m_b/m_B)$. Finally, we have an additional factor of $4$ on the right-hand-side of Eqn. \eqref{K2perpc}. In this way, we are able to recover the expression for $A_I(0)$ as derived in Ref. \cite{Kagan:2001zk}. The numerical impact of these changes is small.  

In the above equations, we have made explicit the scale dependence of the next-to-leading log (NLL) Wilson coeffecients which we take at $\mu_b \sim m_b$ or at the hadronic scale $\mu_h=\sqrt{\Lambda_h \mu_b}$ ($\Lambda_h=0.5~\mbox{GeV}$). We take $m_b=4.6~\mbox{GeV}$ following Ref. \cite{Feldmann:2002iw}. We evolve the Wilson coefficients from the electroweak scale $\mu_W= M_W$ down to the scales $\mu_{b,h}$ using the renormalization group equations. Details of this computation can be found in  Appendix B. The resulting values of the NLL Wilson coefficients at the two scales $\mu=m_b$ and $\mu=\sqrt{\Lambda_h m_b}$ are shown in Table \ref{tab:Wilson}. Note that we use the $3$-loop formula for the running strong coupling $\alpha_s$ (see Appendix A). 

\begin{table}[t]
\centerline{\parbox{14cm}{\caption{\label{tab:Wilson}
NLL Wilson coefficients at the scale $\mu_b=4.6\,$GeV and $\mu_h=1.52$ GeV. Input parameters are 
$\alpha_s(M_z)=0.1184$, $m_t^{\rm pole}=173.5$\,GeV, 
$M_W=80.385$\,GeV and $\sin^2\!\theta_W=0.23$. 
}}}
\vspace{0.1cm}
\begin{center}
\begin{tabular}{|l|c|c|c|c|c|c|}
\hline\hline
\rule[-2mm]{0mm}{7mm}
 & ${\overline{C}}_1$ & ${\overline{C}}_2$ & ${\overline{C}}_3$ & ${\overline{C}}_4$ & ${\overline{C}}_5$
 & ${\overline{C}}_6$ \\
\hline
\rule[-0mm]{0mm}{4mm}
$\mu_b$  & $-0.1482$  & $1.0597$      & $0.0116$ & 
      $-0.0347$ & $0.0099$  & $-0.0393$ \\
\hline
$\mu_h$ & $-0.3423$  & $1.1577$      & $0.0223$ &
      $-0.0629$ & $0.0179$  & $-0.0912$
\\
\hline
\rule[-2mm]{0mm}{7mm}
 & $C_7^{\rm eff}$ & $C_8^{\rm eff}$ & $C_9$ & $C_{10}$
 & &  \\
\hline
\rule[-0mm]{0mm}{4mm}
$\mu_b$  &$-0.3075$ & $-0.1690$ & 
     $4.2381 $ & $-4.6405$
 & & \\
\hline
$\mu_h$ & $-0.3590$ & $-0.2112$ & 
      $4.5019$   & $-4.6405$ 
 & \raisebox{2.5mm}[-2.5mm]{} 
 & \raisebox{2.5mm}[-2.5mm]{} \\
\hline\hline
\end{tabular}
\end{center}
\end{table}

\section{Distribution Amplitudes and Soft Form factors}
We now focus on the non-perturbative inputs namely the Distribution Amplitudes (and decay constants) as well as the soft form factors. 
In Eqns. \eqref{K1perpa}, \eqref{K1perpb} and \eqref{K1perpc}, $\phi_{K^*}^{\perp}(u)$ is the twist-$2$ DA of the transversely polarized $K^*$ while  in Eqns. \eqref{K1parallelb}, \eqref{K1parallelc},  $\phi_{K^*}^{\parallel}(u)$ is the twist-$2$ DA of the longitudinally polarized $K^*$. Note that, to leading twist-$2$ accuracy,  the DAs $g_{K^*}^{\perp(v,a)}(u)$ appearing in Eqn. \eqref{K2perpc} can be expressed in terms of the twist-$2$ DA $\phi_{\parallel}(u)$ \cite{Kagan:2001zk}. In this paper, we shall use the AdS/QCD holographic twist-$2$ DAs which were derived previously in Ref. \cite{Ahmady:2013cva}:
\begin{equation}
\phi_{K^*}^\parallel(z,\mu) =\frac{N_c}{\pi f_{K^*} M_{K^*}} \int \d
r \mu
J_1(\mu r) [M_{K^*}^2 z(1-z) + m_{\bar{q}} m_{s} -\nabla_r^2] \frac{\phi_{K^*}^L(r,z)}{z(1-z)} \;,
\label{phiparallel-phiL}
\end{equation}
\begin{equation}
\phi_{K^*}^\perp(z,\mu) =\frac{N_c }{\pi f_{K*}^{\perp}} \int \d
r \mu
J_1(\mu r) [m_s - z(m_s-m_{\bar{q}})] \frac{\phi_{K^*}^T(r,z)}{z(1-z)} \;,
\label{phiperp-phiT}
\end{equation}
where $\phi_{K^*}^{L,T}(r,z)$ are the holographic wavefunctions obtained by solving the holographic light-front Schroedinger equation \cite{deTeramond:2008ht, deTeramond:2012rt, Brodsky:2013npa}. 
Explicitly \cite{Ahmady:2013cva}
\begin{equation}
\phi_{K^*}^{L,T} (z,\zeta)= \mathcal{N}_{L,T}
\frac{\kappa}{\sqrt{\pi}}\sqrt{z(1-z)} \exp
\left(-\frac{\kappa^2 \zeta^2}{2}\right)
\exp\left \{-\left[\frac{m_s^2-z(m_s^2-m^2_{\bar{q}})}{2\kappa^2 z (1-z)} \right]
\right \} \label{AdS-QCD-wfn-K*}
\end{equation}
with $\kappa=M_{K^*}/\sqrt{2}=0.63$ GeV and where $\zeta=\sqrt{z(1-z)}r$ is the holographic variable which maps onto the fifth dimension in AdS space \cite{deTeramond:2008ht}. In the above equations,  $r$ is the transverse distance between the quark and antiquark and $z$ is the fraction of the meson\rq{}s light-front momentum carried by the quark.

We are also able to compute the decay constants using the holographic wavefunction via the following equations \cite{Ahmady:2013cva}

\begin{equation}
f_{K^*} M_{K^*} = \frac{N_c}{\pi}  \int_0^1 \d z
\left.[z(1-z)M^{2}_{K^*} + m_{\bar{q}} m_{s} -\nabla_{r}^{2}]
\frac{\phi_{K^*}^L(r,z)}{z(1-z)}
\right|_{r=0} \;,
\label{vector-decay}
\end{equation}
and
\begin{equation}
f_{K^*}^{\perp}(\mu) =\frac{N_c}{\pi} \int_0^1 \d z (m_{s} -z(m_s-m_{\bar{q}})) \int \d r \mu J_1(\mu r)  \frac{\phi_{K^*}^T(r,z)}{z(1-z)}
\label{tensor-decay-mu} \;.
\end{equation}
Note that the DAs and the transverse decay constant $f_{K^*}^{\perp}$ are scale-dependent. Here we compute  them at the hadronic scale $\mu=2$ GeV. Using constituent quark masses of $m_{u,d}=0.35$ GeV and $m_s=0.48$ GeV, we obtain $f_{K^*}=0.225 $ GeV and $f_{K^*}^{\perp}(2~\mbox{GeV})= 0.119$ GeV. 

\begin{figure}

\centering     
\subfigure{\includegraphics[width=0.49\textwidth]{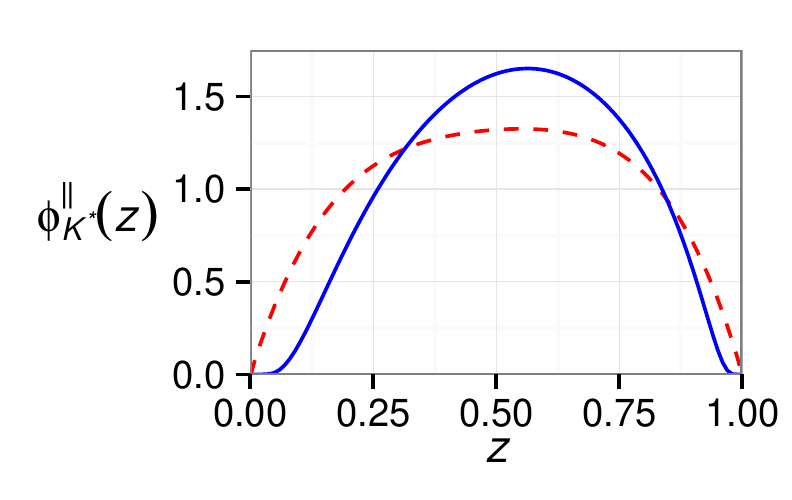}}
\subfigure{\includegraphics[width=0.49\textwidth]{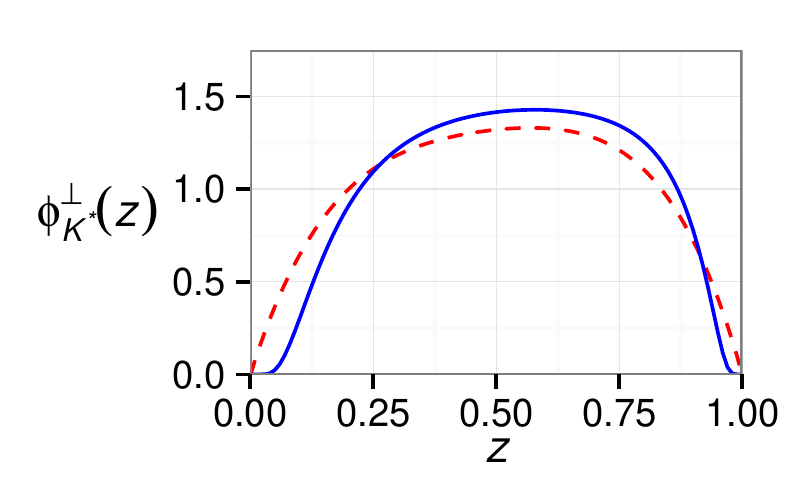}}

\caption{The AdS/QCD DAs (solid blue) compared to the SR DAs (dashed red) at a scale $\mu=2$ GeV.} \label{fig:DAs}
\end{figure}

Sum Rules \cite{Ball:2006eu,Ball:2007zt} are able to  predict the moments of the DAs: 
\begin{equation}
\langle \xi_{\parallel, \perp}^n \rangle_\mu = \int \d z \; \xi^n \phi_{K^*}^{\parallel,\perp} (z, \mu)  
\end{equation}
where $\xi=2z-1$. The first two moments are available in the standard SR approach \cite{Ball:2007zt}. The twist-$2$ DA are then reconstructed as a truncated Gegenbauer expansion
\begin{equation}
\phi_{K^*}^{\parallel,\perp}(z, \mu) = 6 z \bar z \left\{ 1 + \sum_{j=1}^{2}
a_j^{\parallel,\perp} (\mu) C_j^{3/2}(2z-1)\right\}  
\label{phiperp-SR}
\end{equation}
where $C_j^{3/2}$ are the Gegenbauer polynomials and the coefficients $a_j^{\parallel,\perp}(\mu)$ are related to the moments $\langle \xi_{\parallel,\perp}^n \rangle_\mu$ \cite{Choi:2007yu}. These moments and coefficients are determined at a low scale $\mu=1$ GeV and can then  be evolved perturbatively to higher scales \cite{Ball:2007zt}.  As $\mu \to \infty$, they vanish and the DAs take their asymptotic shapes.  We quote the following values from Ref. \cite{Ball:2007zt}: $a_1^{\parallel}(2~\mbox{GeV})=0.02$, $a_2^{\parallel}(2~\mbox{GeV})=0.08$, $a_1^{\perp}(2~\mbox{GeV})=0.03$, $a_2^{\perp}(2~\mbox{GeV})=0.08$ while $f_{K^*}=0.220$ GeV and $f_{K^*}^{\perp}(2~\mbox{GeV})=0.163$ GeV. The AdS/QCD DAs are compared to the SR DAs in Fig. \ref{fig:DAs}. 

As for the soft non-perturbative form factors $\xi_{\perp,\parallel}$ appearing in Eqns. \eqref{Fq2}, \eqref{Gq2}, \eqref{bperpq2} and \eqref{bparallelq2}, we shall compute them in the heavy quark/large recoil limit.  In this limit, the seven $B \to K^*$ transition form factors, which we compute using LCSR with AdS/QCD  DAs  \cite{Ahmady:2014sva}, can be expressed in terms of the two soft form factors: 
\begin{equation}
A_1(E_{K^*})= \frac{2 E_{K^*}}{m_B + m_{K^*}}\xi_{\bot}(E_{K^*}),
\label{xiperpA}
\end{equation}

\begin{equation}
V(E_{K^*})= \frac{m_B+m_{K^*}}{m_B}\xi_{\bot}(E_{K^*}),
\label{xiperpV}
\end{equation}

\begin{equation}
T_1(E_{K^*})=\xi_{\bot}(E_{K^*}),
\label{xiperpT1}
\end{equation}

\begin{equation}
T_2(E_{K^*})=\frac{2 E_K^*}{m_B}\xi_{\bot}(E_{K^*}),
\label{xiperpT2}
\end{equation}

\begin{equation}
A_2(E_{K^*})= \frac{m_B}{m_B-m_{K^*}}
\bigg[\xi_{\bot}(E_{K^*})- \xi_{\|}(E_{K^*})\bigg],
\label{xiperpA2}
\end{equation}

\begin{equation}
A_0(E_{K^*}) = \frac{E_K^*}{m_K^*} \xi_{\|}(E_{K^*}),
\label{xiperpA0}
\end{equation}

\begin{equation}
T_3(E_{K^*})=\xi_{\bot}(E_K^*) - \xi_{\|}(E_{K^*}).
\label{xiperpT3}
\end{equation}
Using Eqs. \eqref{xiperpA}, \eqref{xiperpV}, \eqref{xiperpT1} and \eqref{xiperpT2}, we do a $3$-parameter fit for \mbox{$0 ~\mbox{GeV}^2 \leq q^2 \leq 8 ~\mbox{GeV}^2$} using the parametric form:

\begin{equation}
\xi_{\perp}(q^2)=\frac{\xi_{\perp}(0)}{1-a \hat{s} + b \hat{s}^2}
\hspace{1cm}
\hat{s} = \frac{q^2}{m_B^2}
\end{equation}
We compute all the form factors appearing on the left-hand-side of the above equations using the LCSR given in Ref. \cite{Aliev:1996hb}. In these LCSR, following Ref. \cite{Aliev:1996hb}, we use a Borel parameter $M_B=8~\mbox{GeV}^2$ and a continuum threshold $s_0=36~\mbox{GeV}^2$. Having obtained $\xi_{\perp}$, we are in a position to use Eqs. \eqref{xiperpA2}, \eqref{xiperpA0} and \eqref{xiperpT3} to do a similar fit for $\xi_{\parallel}$. The resulting fitted parameters are shown  in table \ref{tab:xi}.

\begin{table}
\centerline{\parbox{14cm}{\caption{\label{tab:xi}
Fitted parameters for the soft form factors. 
}}}
\vspace{0.1cm}
\begin{center}
\begin{tabular}{|c| c|c|c|}
\hline
\mbox{Model}&a&b&$\xi_\perp(0)$  \\
\hline
\mbox{AdS/QCD}&1.662&0.610&0.245\\
\hline
\mbox{SR}&1.599&0.526&0.283\\
\hline
\end{tabular}
\hspace{1cm}
\begin{tabular}{| c | c| c|c|}
\hline
\mbox{Model}&a&b&$\xi_\parallel(0)$  \\
\hline
\mbox{AdS/QCD}&2.181&1.166&0.076\\
\hline
\mbox{SR}&2.023&0.965&0.076  \\
\hline
\end{tabular}
\end{center}
\end{table}

\begin{figure}
\centering     
\subfigure{\includegraphics[width=0.49\textwidth]{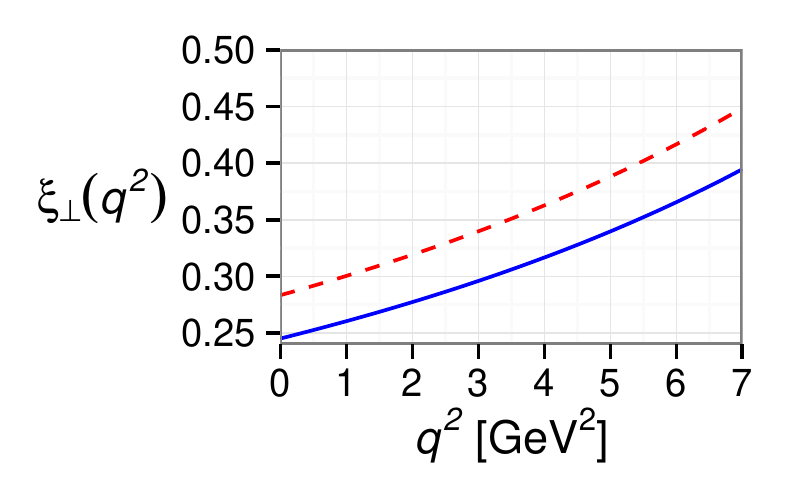}}
\subfigure{\includegraphics[width=0.49\textwidth]{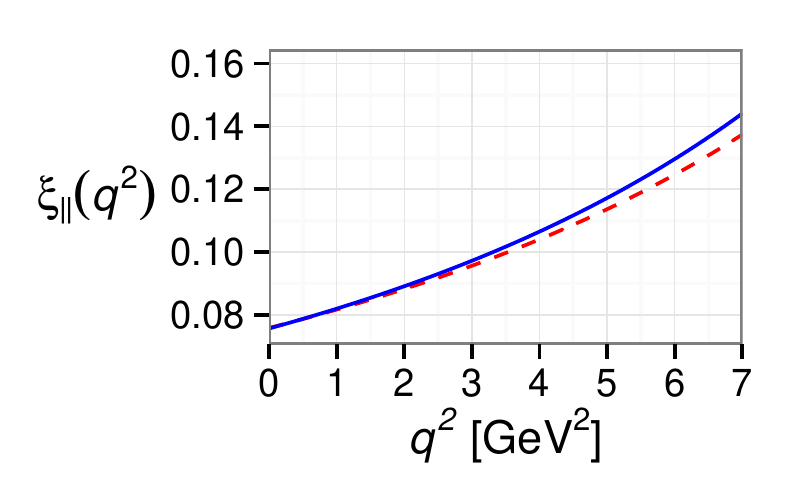}}

\caption{The soft form factors $\xi_{\perp}$ and $\xi_{\parallel}$ as a function of $q^2$. Solid blue: AdS/QCD. Dashed red: Sum Rules.} \label{fig:xi}
\end{figure}

Having specified the DAs and soft form factors, we have now all the ingredients to compute the isospin asymmetry.

\section{Results}
\begin{figure}
\includegraphics[width=1.0\textwidth]{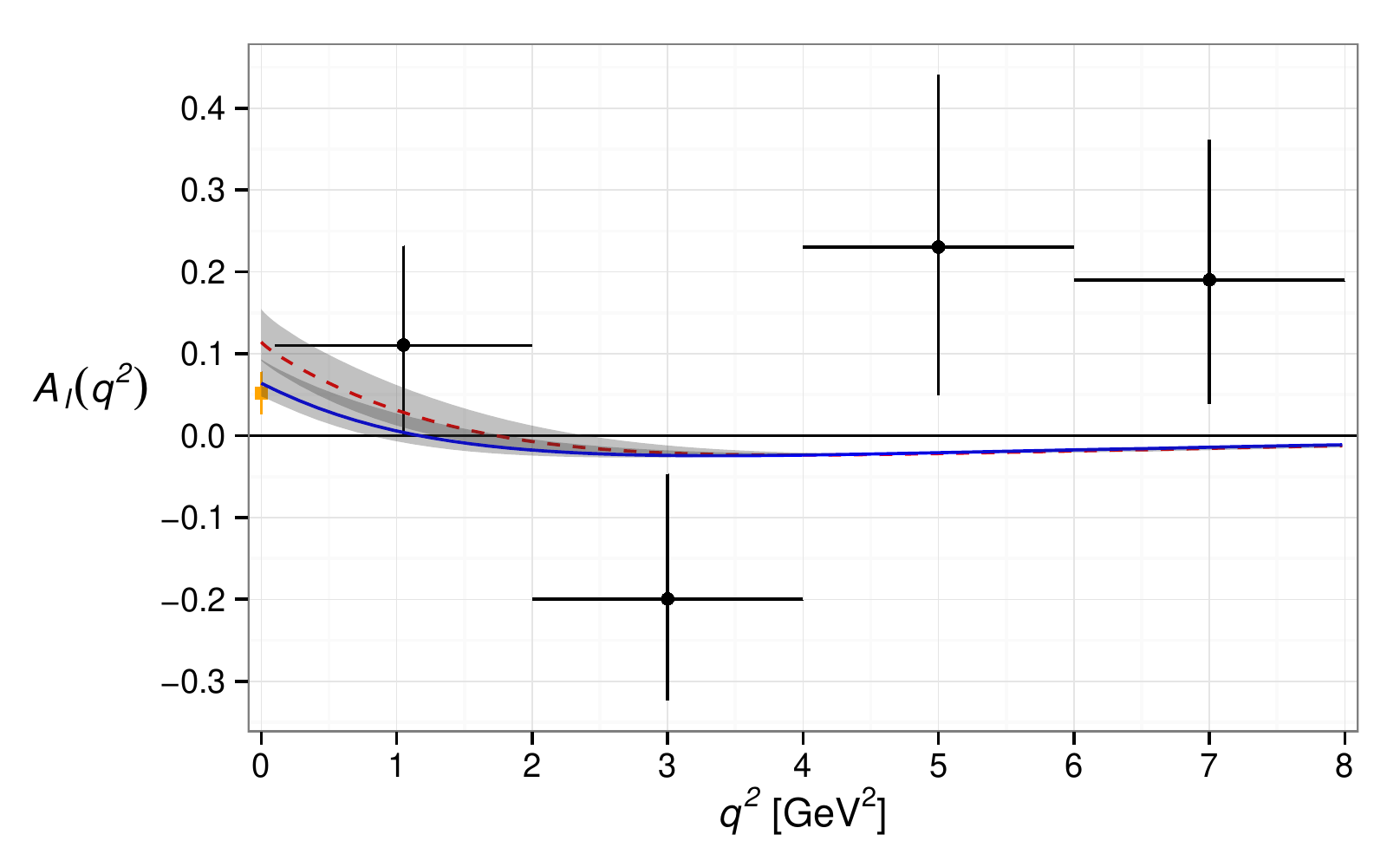} 
\caption{The isospin asymmetry in $B \to K^* \mu^+ \mu^-$ as a function of $q^2$. The AdS/QCD prediction (solid blue) and the Sum Rules prediction (dashed red)  compared to the LHCb data \cite{Aaij:2012cq}. The orange square datum from PDG \cite{Beringer:1900zz} is the isospin asymmetry at $q^2=0$, i.e. for $B \to K^* \gamma$. } \label{fig:IAq2}
\end{figure}

Before discussing our predictions, let us point out that the integral $X_{\perp}(0)$ in Eq. \eqref{K1perpb} diverges at the end-point with the SR DA and that we regulate this divergence using a cut-off as in Ref. \cite{Feldmann:2002iw}. The AdS/QCD and Sum Rules predictions for the isospin asymmetry are shown in Fig. \ref{fig:IAq2}. The uncertainty band for each prediction is obtained by varying the renormalization scale $\mu_b$ between $m_b/2$ and $2m_b$. As can be seen, our predictions are consistent with the LHCb data in the two lowest  $q^2$ bins. At higher $q^2$, we predict a negative isospin asymmetry while the current LHCb data seem to indicate a positive asymmetry. We note that the theoretical computations of Ref. \cite{Feldmann:2002iw} and \cite{Lyon:2013gba} also predict a negative asymmetry in this kinematic region.

\begin{figure}
\includegraphics[width=1.0\textwidth]{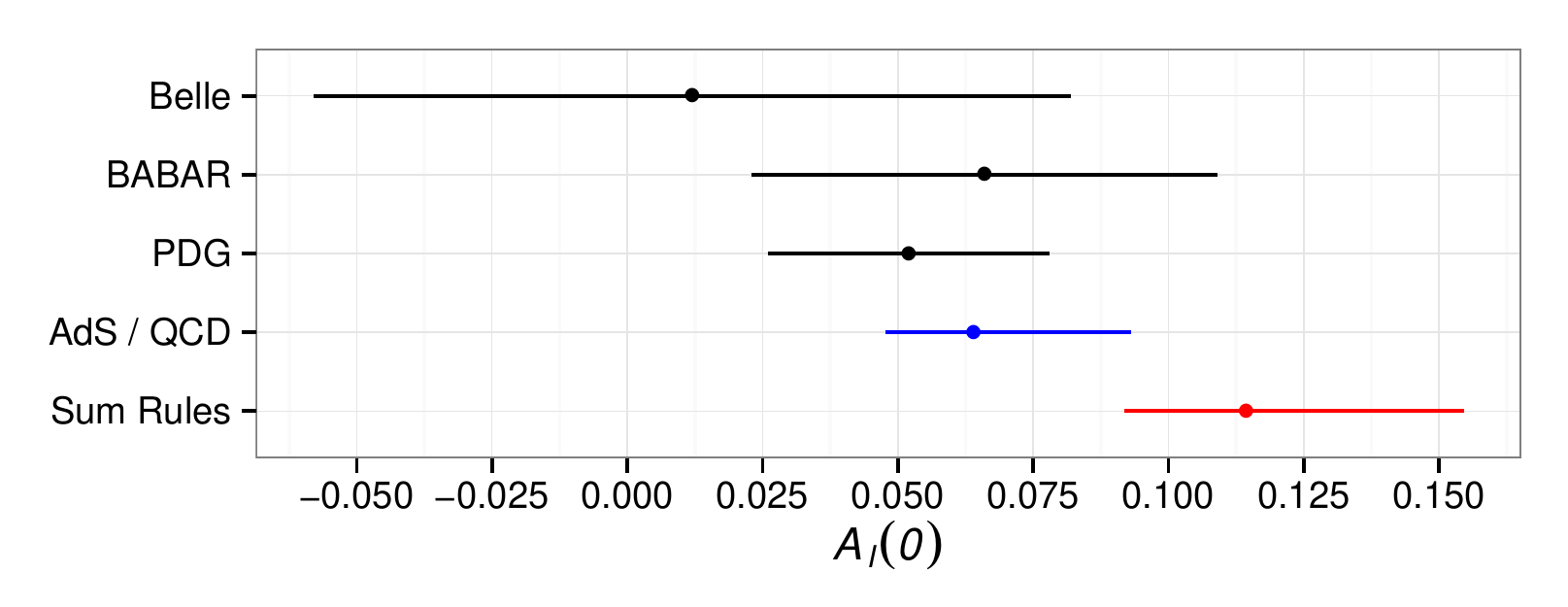} 
\caption{The isospin asymmetry at $q^2=0$, i.e. for $B \to K^* \gamma$. The AdS/QCD prediction (blue) and the Sum Rules (red) predictions compared to the data from Belle \cite{Nakao:2004th}, BaBar \cite{Aubert:2009ak} and PDG \cite{Beringer:1900zz}.} \label{fig:IA0}
\end{figure}

We extrapolate our predictions down to $q^2=0$ in order to update our AdS/QCD prediction \cite{Ahmady:2013cva} of the isospin asymmetry in $B \to K^* \gamma$. We obtain an asymmetry of $6.4\%$ in agreement with the PDG average value of $5.2 \pm 2.6$ \cite{Beringer:1900zz}. Our updated prediction is higher than that ($3.2 \%$) obtained  in Ref. \cite{Ahmady:2013cva} because of different input parameters and a more careful evaluation of the Wilson coefficients at two different scales as explained earlier in this paper. More importantly, we use our AdS/QCD prediction for the form factor $\xi_{\perp}(0) \approx T_1(0)=0.24$ instead of the higher Sum Rules value ($T_1(0)=0.31$ \cite{Ball:2006eu}) used in Ref. \cite{Ahmady:2013cva}. Note that our AdS/QCD prediction for $\xi_{\perp}(0)$ is in very good agreement with the empirical estimate $\xi_{\perp}(0)=0.24 \pm 0.06$ taken from Ref. \cite{Beneke:2001at}. We note that the Sum Rules prediction slightly overshoots the PDG datum at $q^2=0$. We compare our predictions for the asymmetry at $q^2=0$ with the available data in Fig. \ref{fig:IA0}.

\begin{figure}
\includegraphics[width=1.0\textwidth]{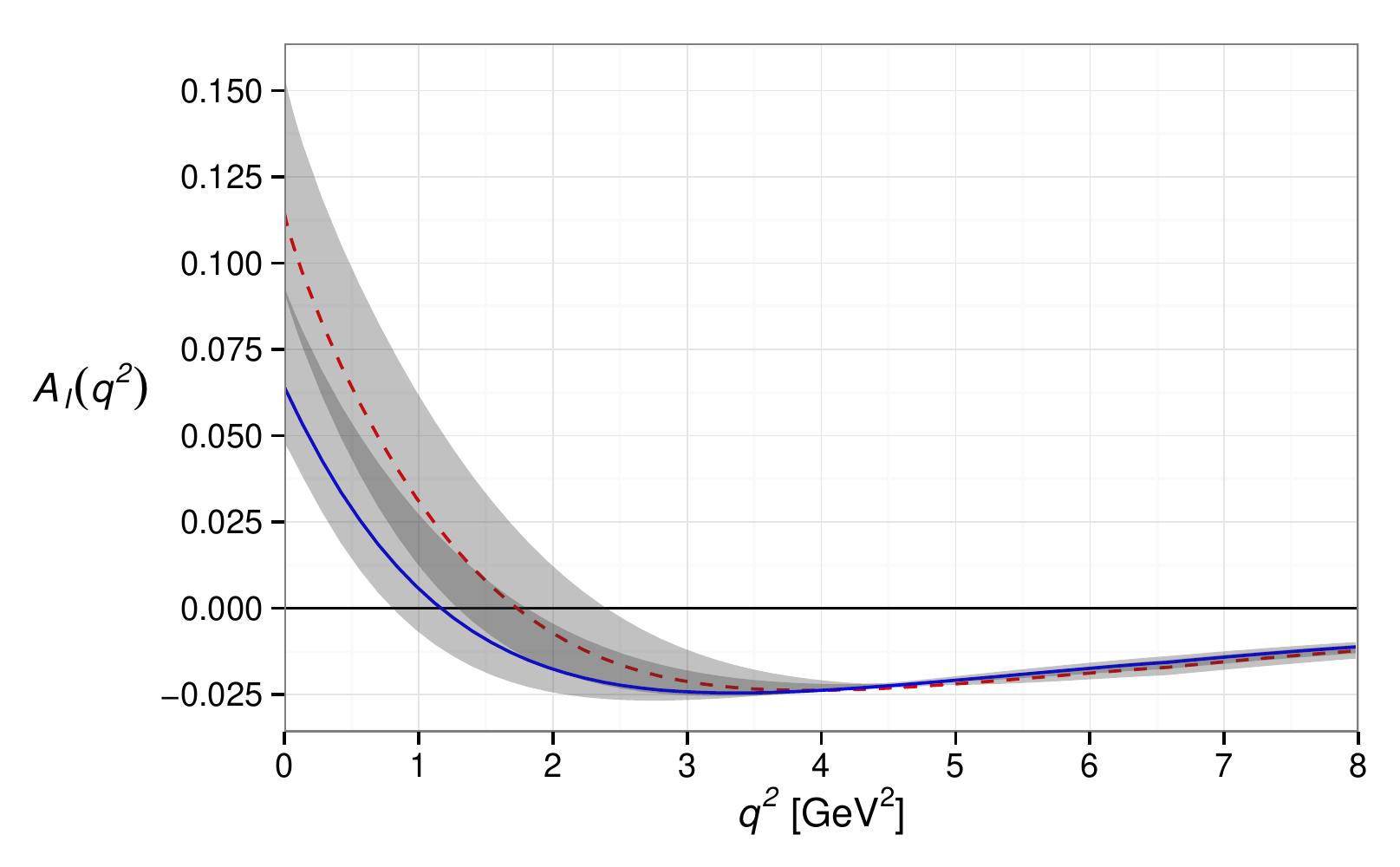} 
\caption{Our predictions for the isospin asymmetry in $B \to K^* \mu^+ \mu^-$ as a function of $q^2$. Red dashed: Sum Rules. Solid blue: AdS/QCD. } \label{fig:IAq2AdSSR}
\end{figure}

In Fig. \ref{fig:IAq2AdSSR}, we take a closer look at AdS/QCD and Sum Rules predictions for the isospin asymmetry distribution. As can be seen, the predictions are distinct as $q^2 \to 0$. Perhaps more interestingly, the predictions become hardly model-dependent and small ($\le 2.5 \%$) for $q^2 \ge 4~\mbox{GeV}^2$. At the same time, they are also hardly sensitive to the variation in the renormalization scale. This means that the isospin asymmetry in this kinematic region is indeed a  clean observable for investigating New Physics signals. Obviously, more precise data would be necessary to reveal any hints of New Physics.

\section{Conclusions}
We have computed the isospin asymmetry in the decay $B \to K^* \mu^+ \mu^-$ using Distribution Amplitudes and decay constants for the $K^*$ as predicted by AdS/QCD and by Sum Rules. Interestingly, the predictions are  hardly model and renormalization scale-dependent in the region of the dimuon mass squared, $q^2 \ge 4~\mbox{GeV}^2$, making more precise measurements of this observable in this kinematic region a  good probe for New Physics.

\section{Acknowledgements}
This research is supported by a team grant from the Natural Sciences and Engineering Research Council of Canada (NSERC).  SL thanks the government of New-Brunswick through SEED-COOP funding. We thank Robyn Campbell for her input in the initial stage of this research. We thank Tracy Lavoie and Taylor Coady for useful discussions, Raymir Mutua  and Liam McManus for reviewing the Appendix. 

\appendix

\section{The strong coupling constant: $\alpha_s(\mu)$}

In this paper, we use the three-loop evolution for $\alpha_s(\mu)$:\cite{Ahmady:2006yr}

\begin{equation}
\alpha_s(\mu) = \frac{4\pi}{\beta_0 \ln(\mu^2 / \Lambda^2)} \left[1 - \frac{\beta_1}{\beta_0^2} \frac{\ln(\ln(\mu^2 / \Lambda^2))}{\ln(\mu^2 / \Lambda^2)} + \frac{\beta_1^2}{\beta_0^4 \ln^2(\mu^2 / \Lambda^2)} \left( \left( \ln(\ln(\mu^2 / \Lambda^2)) - \frac{1}{2} \right)^2 + \frac{\beta_2 \beta_0}{2 \beta_1^2} -\frac{5}{4} \right) \right]
\end{equation}

where 

\begin{equation}
\beta_0 = \frac{33 - 2n_f}{3} \hspace{1cm} \beta_1 = \frac{306 - 38n_f}{3} \hspace{1cm} \beta_2 = 2857 - \frac{5033}{9}n_f + \frac{325}{27}n_f^2
\end{equation}

and where  $n_f$ is the number of active flavors according to which the value of $\Lambda$ is fixed using threshold matching conditions.

\section{Wilson Coefficients}
We start by computing the \lq\lq{}barred\rq\rq{} Wilson coefficients $\overline{C}_{1-6}$ used in this paper. The ``barred"  Wilson coefficients $\overline{C}_{1-6}$ are defined in the basis used in \cite{Beneke:2001at}. By construction, at leading log (LL) accuracy they coincide  with the Wilson coefficients $C_{1-6}$ of the standard basis \cite{Buras:1998raa}. At next-to-leading log (NLL) accuracy, the two sets of coefficients are related by the equations \cite{Beneke:2001at}
\begin{equation}
\overline{C}_i (\mu) = C_i(\mu) + \frac{\alpha_s(\mu)}{4\pi} T_{ij} C_j(\mu) \hspace{1cm} i \in \{1, 2, 3, 4, 5, 6\}
\end{equation}
where
\begin{equation}
T_{ij} = 
\left(
\begin{array}{cccccc}
\frac{7}{3} & 2 & 0 & 0 & 0 & 0 \\
1 & -\frac{2}{3} & 0 & 0 & 0 & 0 \\
0 & 0 & -\frac{178}{27} & -\frac{4}{9} & \frac{160}{27} & \frac{13}{9} \\
0 & 0 & \frac{34}{9} & \frac{20}{3} & -\frac{16}{9} & -\frac{13}{3} \\
0 & 0 & \frac{164}{27} & \frac{23}{9} & -\frac{146}{27} & -\frac{32}{9} \\
0 & 0 & -\frac{20}{9} & -\frac{23}{3} & \frac{2}{9} & \frac{16}{3}
\end{array}
\right)
\end{equation}
The NLL coefficients $C_{1-6}$ are themselves given by \cite{Buras:1998raa}:
\begin{equation}
C_i(\mu) = C_i^0(\mu) + \frac{\alpha_s(\mu)}{4\pi} C_i^1(\mu)
\label{gWilson}
\end{equation}
where  
\begin{equation}
C_i^0(\mu) = \sum_{j = 1}^8 k_{ij}\eta^{a_{j}} \hspace{1cm} C_i^1(\mu) = \sum_{j = 1}^8(e_{ij}\eta E_0(x_t) + f_{ij} + g_{ij}\eta)\eta^{a_j}
\label{LLWC}
\end{equation}
with
\begin{equation}
\eta = \frac{\alpha_s(m_W)}{\alpha_s(\mu)} \hspace{1cm} E_0(x) = \frac{x(18 - 11x - x^2)}{12(1-x)^3} + \frac{x^2(15 - 16x + 4x^2)}{6(1-x)^4}\ln(x) - \frac{2}{3}\ln(x)
\end{equation}
and\cite{Ahmady:2006yr}
\begin{equation}
x_t = \frac{\overline{m_t}(m_W)^2}{m_W^2} \;.
\label{xt}
\end{equation}
In Eq. \eqref{xt}, the running top mass  is given by 
\begin{equation}
\overline{m_t}(\mu) = \overline{m_t}(m_t)\left( \frac{\alpha_s(\mu)}{\alpha_s(m_t)} \right)^\frac{\gamma_0^m}{2\beta_0} \left(1 + \frac{\alpha_s(m_t)}{4\pi}  \frac{\gamma_0^m}{2 \beta_0}  \left( \frac{\gamma_1^m}{\gamma_0^m} - \frac{\beta_1}{\beta_0} \right) \left( \frac{\alpha_s(\mu)}{\alpha_s(m_t)} - 1\right)\right)
\end{equation}
where ($m_t$ is the pole mass of the top quark)
\begin{equation}
\overline{m_t}(m_t) = m_t\left( 1- \frac{4}{3} \frac{\alpha_s(m_t)}{\pi}\right) \hspace{1cm} \gamma_0^m = 8 \hspace{1cm} \gamma_1^m = \frac{404}{3} -\frac{40}{9}n_f
\end{equation}
To completely specify \eqref{LLWC}, we also need the following matrices:

\begin{equation}
k_{ij} = 
\left(
\begin{array}{cccccccc}
0 & 0 & \frac{1}{2} & -\frac{1}{2} & 0 & 0 & 0 & 0 \\
0 & 0 & \frac{1}{2} & \frac{1}{2} & 0 & 0 & 0 & 0 \\
0 & 0 & -\frac{1}{14} & \frac{1}{6} & 0.0510 & -0.1403 & -0.0113 & 0.0054 \\
0 & 0 & -\frac{1}{14} & -\frac{1}{6} & 0.0984 & 0.1214 & 0.0156 & 0.0026 \\
0 & 0 & 0 & 0 & -0.0397 & 0.0117 & -0.0025 & 0.0304 \\
0 & 0 & 0 & 0 & 0.0335 & 0.0239 & -0.0462 & -0.0112
\end{array}
\right)
\end{equation}

\begin{equation}
e_{ij} = 
\left(
\begin{array}{cccccccc}
0 & 0 & 0 & 0 & 0 & 0 & 0 & 0 \\
0 & 0 & 0 & 0 & 0 & 0 & 0 & 0 \\
0 & 0 & 0 & 0 & 0.1494 & -0.3726 & 0.0738 & -0.0173 \\
0 & 0 & 0 & 0 & 0.2885 & 0.3224 & -0.1025 & -0.0084 \\
0 & 0 & 0 & 0 & -0.1163 & 0.0310 & 0.0162 & -0.0975 \\
0 & 0 & 0 & 0 & 0.0982 & 0.0634 & 0.3026 & 0.0358
\end{array}
\right)
\end{equation}

\begin{equation}
f_{ij} = 
\left(
\begin{array}{cccccccc}
0 & 0 & 0.8136 & 0.7142 & 0 & 0 & 0 & 0 \\
0 & 0 & 0.8136 & -0.7142 & 0 & 0 & 0 & 0 \\
0 & 0 & -0.0766 & -0.1455 & -0.8848 & 0.4137 & -0.0114 & 0.1722 \\
0 & 0 & -0.2353 & -0.0397 & 0.4920 & -0.2758 & 0.0019 & -0.1449 \\
0 & 0 & 0.0397 & 0.0926 & 0.7342 & -0.1262 & -0.1209 & -0.1085 \\
0 & 0 & -0.1191 & -0.2778 & -0.5544 & 0.1915 & -0.2744 & 0.3568
\end{array}
\right)
\end{equation}

\begin{equation}
g_{ij} = 
\left(
\begin{array}{cccccccc}
0 & 0 & 1.0197 & 2.9524 & 0 & 0 & 0 & 0 \\
0 & 0 & 1.0197 & -2.9524 & 0 & 0 & 0 & 0 \\
0 & 0 & -0.1457 & -0.9841 & 0.2303 & 1.4672 & 0.0971 & -0.0213 \\
0 & 0 & -0.1457 & 0.9841 & 0.4447 & -1.2696 & -0.1349 & -0.0104 \\
0 & 0 & 0 & 0 & -0.1792 & -0.1221 & 0.0213 & -0.1197 \\
0 & 0 & 0 & 0 & 0.1513 & -0.2497 & 0.3983 & 0.0440
\end{array}
\right)
\end{equation}
and \cite{Buchalla:1995vs}
\begin{equation}
a_i = 
\left(
\begin{array}{cccccccc}
\frac{14}{23} & {16}{23} & \frac{6}{23} & -\frac{12}{23} & 0.4086 & -0.4230 & -0.8994 & 0.1456
\end{array}
\right)
\end{equation}

We now turn to the computation of the Wilson coefficients $C_{7,8}^{\rm eff}$.  At NLL accuracy, they are given by
\begin{equation}
C_{7,8}^{\rm eff}(\mu) = C_{7,8}^{(0) \rm eff}(\mu) + \frac{\alpha_s(\mu)}{4\pi} C_{7,8}^{(1) \rm eff}(\mu)
\label{gWilson78}
\end{equation}
where \cite{Buras:1998raa} 
\begin{equation}
C_7^{(0) \rm eff}(\mu) = \eta^{\frac{16}{23}} C_7^{(0)}(m_W) + \frac{8}{3}\left( \eta^{\frac{14}{23}} - \eta^{\frac{16}{23}} \right) C_8^{(0)}(m_W) + C_2^{(0)}(m_W) \sum_{i=1}^{8}h_i\eta^{a_i} \;,
\end{equation}
and
\begin{multline}
C_7^{(1) \rm eff}(\mu) = \eta^{\frac{39}{23}} C_7^{(1) \rm eff}(m_W) + \frac{8}{3} \left(\eta^{\frac{37}{23}} - \eta^{\frac{39}{23}} \right) C_8^{(1) \rm eff}(m_W) \\ + \left( \frac{297664}{14283} \eta^{\frac{16}{23}} - \frac{7164416}{357075}\eta^{\frac{14}{23}} + \frac{256868}{14283}\eta^{\frac{37}{23}} - \frac{6698884}{357075} \eta^\frac{39}{23} \right) C_8^{(0)}(m_W) \\ + \frac{37208}{4761} \left( \eta^\frac{39}{23} - \eta^\frac{16}{23} \right) C_7^{(0)}(m_W) + \sum_{i=1}^8(e'_i \eta E_0(x_t) + f'_i + g'_i\eta)\eta^{a_i} \;,
\end{multline}
while
\begin{equation}
C_8^{(0) \rm eff}(\mu) = \eta^{\frac{14}{23}}C_8^{(0)}(m_W)+C_2^{(0)}(m_W)\sum_{i=1}^8 \overline{h}_i \eta^{a_{i}} \;.
\end{equation}
and \cite{Greub:2000sy}
\begin{multline}
C_8^{(1) \rm eff}(\mu) = \eta^{\frac{37}{23}} C_8^{(1) \rm eff}(m_W) + 6.7441 \left( \eta^{\frac{37}{23}} - \eta^{\frac{14}{23}} \right) C_8^{(0)}(m_W) \\ + \sum_{i=1}^8 \left( \overline{e'}_i \eta C_4^{(1) \rm eff}(m_W) + \left( \overline{f'}_i + \overline{g'}_i \eta \right) C_2^{(0)}(m_W) + \overline{l'}_i \eta C_1^{(1) \rm eff}(m_W) \right) \eta^{a_1} \;.
\end{multline}

The various Wilson coefficient at a scale $m_W$ are given below: \cite{Greub:2000sy}

\begin{equation}
C_1^{(1) \rm eff}(m_W) = 15 \hspace{1cm} C_2^{(0)}(m_W) = 1 \hspace{1cm} C_4^{(1) \rm eff}(m_W) = E_0(x_t) -\frac{2}{3} 
\end{equation}
while\cite{Buras:1998raa}
\begin{equation}
C_7^{(0)}(m_W) = \frac{3x_t^3 - 2x_t^2}{4(x_t-1)^4}\ln(x_t) + \frac{-8x_t^3-5x_t^2+7x_t}{24(x_t-1)^3}
\end{equation}

\begin{multline}
C_7^{(1) \rm eff}(m_W) = \frac{-16x_t^4 -122x_t^3 + 80x_t^2 -8x_t}{9(x_t-1)^4}\Li_2\left(1 - \frac{1}{x_t} \right) + \frac{6x_t^4 + 46x_t^3 - 28x_t^2}{3(x_t - 1)^5}\ln^2x_t \\ + \frac{-102x_t^5 -588x_t^4 -2262x_t^3 + 3244x_t^2 - 1364x_t +208}{81(x_t - 1)^5}\ln(x_t) \\ + \frac{1646x_t^4 + 12205x_t^3 - 10740x_t^2 + 2509x_t -436}{486(x_t -1)^4}
\end{multline}

\begin{equation}
C_8^{(0)}(m_W) = \frac{-3x_t^2}{4 (x_t - 1)^4}\ln(x_t) + \frac{-x_t^3 + 5x_t^2+2x_t}{8( x_t -1)^3}
\end{equation}

\begin{multline}
C_8^{(1) \rm eff}(m_W) = \frac{-4x_t^4 + 40x_t^3 + 41x_t^2 + x_t}{6( x_t -1)^4}\Li_2 \left(1 - \frac{1}{x_t} \right) + \frac{-17x_t^3 - 31x_t^2}{2( x_t - 1)^5}\ln^2(x_t) \\ + \frac{-210x_t^5 + 1086x_t^4 + 4893x_t^3 + 2857x_t^2 - 1994x_t + 280}{216( x_t - 1)^5}\ln(x_t) \\ + \frac{737x_t^4 - 14102x_t^3 - 28209x_t^2 + 610x_t - 508}{1296( x_t -1 )^4}
\end{multline}
where $\Li_2(z)$ is Spence's function defined by

\begin{equation}
\Li_2(z) = - \int_0^z \frac{\ln(1 - u)}{u} du
\end{equation} 


{
\begin{table}

\begin{tabular}{| c | c c c c c c c c |}
\hline
i & 1 & 2 & 3 & 4 & 5 & 6 & 7 & 8\\
\hline
$e'$ & $\frac{4661194}{816831}$ & $-\frac{8516}{2217}$ & 0 & 0 & -1.9043 & -0.1008 & 0.1216 & 0.0183 \\
$f'$ & -17.3023 & 8.5027 & 4.5508 & 0.7519 & 2.0040 & 0.7476 & -0.5385 & 0.0914 \\
$g'$ & 14.8088 & -10.8090 & -0.8740 & 0.4218 & -2.9347 & 0.3971 & 0.1600 & 0.0225 \\
$h$ & 2.2996 & -1.0880 & $-\frac{3}{7}$ & $-\frac{1}{14}$ & -0.6494 & -0.0380 & -0.0185 & -0.0057 \\
$\overline{e'}$ & 2.1399 & 0 & 0 & 0 & -2.6788 & 0.2318 & 0.3741 & -0.0670 \\
$\overline{f'}$ & -5.8157 & 0 & 1.4062 & -3.9895 & 3.2850 & 3.6851 & -0.1424 & 0.6492 \\
$\overline{g'}$ & 3.7264 & 0 & 0 & 0 & -3.2247 & 0.3359 & 0.3812 & -0.2968 \\
$\overline{l'}$ & 0.2169 & 0 & 0 & 0 & -0.1793 & -0.0730 & 0.0240 & 0.0113 \\
$\overline{h}$ & 0.8623 & 0 & 0 & 0 & -0.9135 & 0.0873 & -0.0571 & 0.0209\\
\hline
\end{tabular}

\caption{Useful numbers in the calculation of $C_7^{\rm eff}$ and $C_8^{\rm eff}$}

\end{table}
}
Finally, we compute the Wilson coefficients $C_9$ and $C_{10}$ at NLL using \cite{Buchalla:1995vs}

\begin{equation}
C_9(\mu) = P_0+\frac{Y_0(x_t)}{\sin^2(\theta_W)} - 4Z_0(x_t) + P_E E_0(x_t)
\end{equation}
where $\theta_W$ is  the Weinberg angle (i.e. $\sin^2(\theta_W) = 0.23$) and the functions $P_0$, $P_E$, $Y_0$ and $Z_0$ are given by:
\begin{equation}
P_0 = \frac{\pi}{\alpha_s(m_W)}(-0.1875 + \sum_{i = 1}^8 p_i \eta^{a_i+1}) + 1.2468 + \sum_{i=1}^8\eta^{a_i}(r_i + s_i \eta)
\end{equation}
\begin{equation}
P_E = 0.1405 + \sum_{i=1}^8 q_i \eta^{a_i+1}
\end{equation}
\begin{equation}
Y_0(x) = C_0(x) - B_0(x) \hspace{1cm} Z_0 = C_0 + \frac{1}{4}D_0 (x)
\end{equation}
where the functions $B_0$, $C_0$ and $D_0$ are defined by:
\begin{equation}
B_0(x) = \frac{1}{4} \left( \frac{x}{1-x} + \frac{x \ln(x)}{(x-1)^2} \right) \hspace{1cm} C_0(x) = \frac{x}{8} \left( \frac{x-6}{x-1} + \frac{3x + 2}{(x-1)^2}\ln(x) \right)
\end{equation}
\begin{equation}
D_0(x) = -\frac{4}{9}\ln(x) + \frac{-19x^3 + 25x^2}{36(x-1)^3} + \frac{x^2(5x^2 - 2x -6)}{18(x-1)^4}\ln(x)
\end{equation}

The $C_{10}$ Wilson coefficient has no scale dependence\cite{Buchalla:1995vs}:
\begin{equation}
C_{10}(\mu) = C_{10}(m_W) = -\frac{Y_0(x_t)}{\sin^2(\theta_W)} \;.
\end{equation}

{
\begin{table}

\begin{tabular}{| c | c c c c c c c c |}
\hline
i & 1 & 2 & 3 & 4 & 5 & 6 & 7 & 8 \\
\hline
$p$ & 0 & 0 & $-\frac{80}{203}$ & $\frac{8}{33}$ & 0.0433 & 0.1384 & 0.1648 & -0.0073\\
$r$ & 0 & 0 & 0.8966 & -0.1960 & -0.2011 & 0.1328 & -0.0292 & -0.1858\\
$s$ & 0 & 0 & -0.2009 & -0.3579 & 0.0490 & -0.3616 & -0.3554 & 0.0072\\
$q$ & 0 & 0 & 0 & 0 & 0.0318 & 0.0918 & -0.2700 & 0.0059\\
\hline
\end{tabular}
\caption{Useful numbers in the calculation of $C_9$.}
\end{table}
}


\bibliographystyle{apsrev}
\bibliography{kstariso}

\end{document}